# Intelligent optimization of mine environmental damage assessment and repair strategies based on deep learning

Qishuo Cheng
University of Chicago, IL, Chicago, USA
qishuoc@uchicago.edu

*Abstract*—In recent decades, financial quantification has emerged and matured rapidly. For financial institutions such as funds, investment institutions are increasingly dissatisfied with the situation of passively constructing investment portfolios with average market returns, and are paying more and more attention to active quantitative strategy investment portfolios. This requires the introduction of active stock investment fund management models. Currently, in my country's stock fund investment market, there are many active quantitative investment strategies, and the algorithms used vary widely, such as SVM, random forest, RNN recurrent memory network, etc. This article focuses on this trend, using the emerging LSTM-GRU gate-controlled long short-term memory network model in the field of financial stock investment as a basis to build a set of active investment stock strategies, and combining it with SVM, which has been widely used in the field of quantitative stock investment. Comparing models such as RNN, theoretically speaking, compared to SVM that simply relies on kernel functions for high-order mapping and classification of data, neural network algorithms such as RNN and LSTM-GRU have better principles and are more suitable for processing financial stock data. Then, through multiple By comparison, it was finally found that the LSTM-GRU gate-controlled long short-term memory network has a better accuracy. By selecting the LSTM-GRU algorithm to construct a trading strategy based on the Shanghai and Shenzhen 300 Index constituent stocks, the parameters were adjusted and the neural layer connection was adjusted. Finally, It has significantly outperformed the benchmark index CSI 300 over the long term. The conclusion of this article is that the research results can provide certain quantitative strategy references for financial institutions to construct active stock investment portfolios.*

## I. INTRODUCTION

The operating logic behind active stock investment funds is significantly different from traditional passive stock investment funds. Relying on actively constructed stock selection or timing or a combination of both strategies, active stock investment funds can obtain significantly higher returns than traditional ones. The income of passive funds relies on mature algorithm combinations. Active funds can expand their income advantages in the medium and long term. While achieving greater returns, active funds can also better reduce the overall risk fluctuations of their investment portfolios. Taking the HS300 index as an example, during the 2015 stock market crash, the index's maximum decline even reached 40%, which would cause a devastating blow to passive stock investment funds [1,2]. China's financial market is developing rapidly, and the stock market is also rapidly upgrading and developing. The connection between the two is getting closer. To improve the investment level of institutional investors in the entire market, a long-term institutional investor investment mechanism is increasingly critical to the stability of the market. . This article focuses on constructing a medium- and long-term quantitative investment strategy, hoping to provide a reference for active stock investment funds to conduct medium- and long-term active quantitative investments. This article combines the latest machine learning algorithm LSTM-GRU gate-controlled long short-term memory network [3-5], and optimizes the construction of composite factors based on the excessive short-selling effect of most A-share factors. According to the Shanghai and Shenzhen 300 constituent stocks Weekly data builds a stock pool, and through this strategy, we find stocks with investment value, build a stock investment portfolio, and achieve a portfolio that performs better than the CSI 300.

The multi-factor stock selection model is one of the classic stock selection models. Now we want to use the improvement of computer computing power and machine learning algorithms to improve its stock selection capabilities. This article is built based on the LSTM-GRU gate-controlled long short-term memory network, and considers the value of messages in different time periods to improve the model parameters. The advantages of this construction are as follows:

Utilize the advantages of neural network models in data modeling. Neural network is a rapidly developing machine learning method. Its principle was first born from the Turing test proposed by Turing. Previously, due to limitations of computing power and data volume, neural networks have been making slow progress. Nowadays, the rapid development of computer technology has solved the above limitations. Modern neural networks (ANN) have developed rapidly, and then developed convolutional neural networks (CNN) focusing on image feature extraction and recurrent memory networks focusing on time series data feature extraction. (RNN) [6] and other nearly a hundred types of neural networks. Recurrent memory networks as the main body and various deformed networks have been widely used in financial data processing. Due to its large data volume and greatly improved modeling accuracy compared with traditional measurement modeling, it has been widely praised.

Avoid the influence of subjective emotions of institutional investors. One of the great benefits of using machine learning for modeling is to avoid the influence of investor sentiment. As we all know, the Chinese stock market is a highly irrational market. Institutional investors also have a "herding effect" and are susceptible to chasing ups and downs. Influenced by factors, machine learning performs parameter tuning based on matrices and is not affected by emotions. This method is more objective and avoids investment bias caused by blind obedience.

Improved the gradient disappearance and gradient explosion problems of RNN recurrent memory network. The RNN recurrent memory network is relatively reliable when processing short-term data information, but when the timeline is extended, RNN often has the problem of "forgetting" the early training weights. The gate-controlled long short-term memory network improves this problem. The reason is that In the calculation of the hidden layer state index $ht$, the gradient is expressed in the form of continuous multiplication. The LSTM long short-term memory network adds a cell state $ct$ and three "control valves" to complete the classification, screening and forgetting of information (essentially weight matrix update) , the control valve uses sigmoid activation function and pointwise multiplication to complete the weight matrix update work. The multi-layer control valve allows the LSTM long short-term memory network to retain certain weights when processing data with a long time span. The GRU gate control loop model is similar to the LSTM model. In fact, the GRU model merges and reconstructs the input and forgetting in the LSTM model control valve into an update gate, and retains the output gate. This can make up for the performance of the LSTM model in some data sets and significantly improve the performance of the LSTM model. significantly improve the time complexity.

## II. FUNDAMENTAL PRINCIPLES

### A. Long short term memory network

The LSTM-GRU gated long short-term memory network has developed rapidly in the field of NLP (natural language processing), but its application in finance, especially the stock field, is in the ascendant. There are currently literatures that mainly use a single LSTM model to predict the overall market trend of the stock market. Currently, there is no Academic literature that can be referred to for real stock investment. This article will apply the LSTM-GRU network to real stock investment, and construct it in a large number of stock pools and within a medium and long-term span to test the stock selection reliability of this algorithm. It is of research value to apply the LSTM-GRU gated long short-term memory network algorithm in the field of stock investment. In order to better establish the model, a large number of past literature and related materials in the field of stock investment are referred to, for example, in order to better study factors Value, referring to the China A-share market quantitative factor white paper and relevant brokerage reports and other documents, selected multiple types of factors with high timeliness as the basic factor library, and combined with the LSTM-GRU gate-controlled long short-term memory network to establish a complete set of quantitative investment strategies . There are also some highlights of the quantitative investment strategy of this article: the construction of targeted composite factors based on the excessive short-selling effect of some factors, the comparison of related machine learning algorithm models and the parameter tuning of the LSTM-GRU gate control long short-term memory network, etc[7].

Based on the LSTM-GRU gate-controlled long short-term memory network algorithm, this paper constructs a stock selection strategy and obtains higher excess returns than the benchmark index CSI 300. Combined with the analysis of financial factors in China's stock market and related research results, an improvement based on the weakening of the short effect of factors has been constructed, and the rate of return and model stability have been improved to a certain extent.

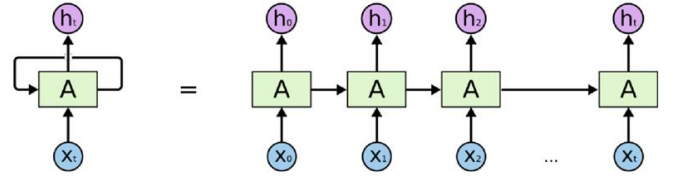

Fig.1 RNN loop operation expansion diagram.

This figure is a simple RNN diagram containing a loop. After a certain neuron A receives data X, it is not calculated only once. According to the characteristics of the RNN cyclic memory network, after the initial data The new input data is continuously circulated within the time range [0,t]. This characteristic gives RNN the ability and characteristics of quantitative analysis of stock investment in the financial field.

### B. Gated loop unit

According to the principle of LSTM, it can be found that the forget gate and the input gate play an important role. GRU combines forget gates and input gates into "update gates" to simplify the information path inside the unit. In addition, it removes the unit state design and adds "reset gates" and other changes. While ensuring that the capabilities of the neural network do not decrease, these changes speed up the calculation and achieve higher efficiency.

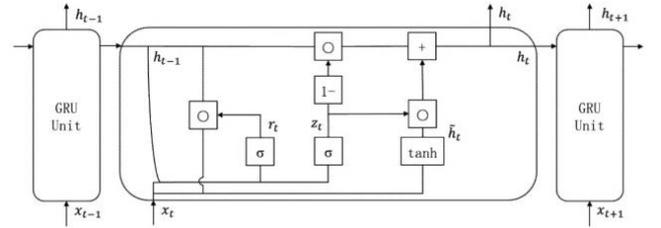

Fig.2 GRU structure diagram.

The GRU structure is shown in Figure 2, where $r_t$ and $z_t$ are reset gates and update gates. The reset gate determines how to combine the new input information $x_t$ with the information of the previous unit $h_{t-1}$. The calculation formula is as follows:

$$r_t = \sigma(W_r \times [h_{t-1}, x_t] + b_r) \# (1)$$

The update gate controls the abandonment and retention of the previous unit information $h_{t-1}$ and the new input information $x_t$. The calculation formula is as follows:

$$z_t = \sigma(W_z \times [h_{t-1}, x_t] + b_z) \# (1)$$

CRU can use only one update gate to play the role of two gates (forgetting gate and input gate), relying on the exquisite use of $z_t$. The relevant formula is as follows:

$$\widetilde{h}_t = tanh(W \times [r_t \circ h_{t-1}, x_t] + b) (1)$$
$$h_t = (1 - z_t) \circ h_{t-1} + z_t \circ \widetilde{h}_t \# (1)$$

where $\widetilde{h}_t$ represents the candidate output, which together with the previous unit information $h_{t-1}$, produces the final information output $h_t$ of the entire unit. In the above formula, $W_r$, $W_z$, $W$, represent the weight matrix, $b_r$, $b_z$ and $b$ represent the bias vector, ° represents the Hadamard product of the matrix, σ and tanh represent the Sigmoid function and the hyperbolic tangent function respectively.

## C. Trading straregy

Trading investment strategy is another common research direction in quantitative stock selection [8]. Kelly was the first to conduct research on constructing dynamic investment portfolios. He proposed to adjust dynamic investment portfolios based on maximizing the expected logarithmic return index as an indicator. This idea of cyclical portfolio adjustment that strives to maximize the target rate of return is very suitable as a quantitative strategy for stock investment. Zhang Weiguo used the distribution mean of the historical return rate of the investment portfolio as the proportional coefficient. Based on this, he built a regression model for the core of the algorithm calculation and achieved good historical backtest results. Wang Jianfeng proposed that we should pay attention to the impact of parameter setting in machine learning on investment, and proposed an improved SVM parameter tuning method to obtain better returns. Meng Xuejing constructed public opinion information as an algorithm consideration, crawled daily public opinion information through a crawler, extracted text, and then used it as an investment evaluation system, achieving certain benefits. Zhao Chen added the financial risks of listed companies as an important consideration into the modeling of the BP neural network, and achieved certain portfolio returns by using weights to strengthen the negative feedback learning of financial risks.

As for the LSTM model, most of the current research results focus on exploring the overall trend of the index and the construction of small portfolio models in the short term, and the number of factor pools in the portfolio is small, usually no more than 30, to test the underlying Stock portfolios lack the persuasiveness of practical stock investment quantitative strategies. In addition, neural network modeling requires parameter adjustment. Currently, there is little literature on LSTM model structure construction and parameter tuning. Some literature is based on the bidirectional LSTM model, that is, building a reverse LSTM model, which improves a certain accuracy, but does not introduce the optimization problem based on the efficiency of the loss function.

## III. NEURAL NETWORK APPLICATIONS

### A. Long short-term memory network model (LSTM)

According to the research of many scholars, LSTM is dedicated to solving the "gradient disappearance" problem of RNN. In 1997, Hochreaer et al. proposed this network to solve the hidden layer calculation problem of RNN.

For the RNN model, in the process of minimizing the loss, we need to derive the loss function and calculate the gradient, because the gradient direction, also called the derivation direction, is the direction in which the value of the loss function decreases fastest. However, when calculating the gradient, the vanishing gradient phenomenon is likely to occur. That is, in the process of finding the minimum value, the gradient function will quickly disappear. The result is that the loss function takes almost infinite time to reach its minimum value. Despite their simplicity, traditional methods are still effective in solving the problem of vanishing gradients. At the same time, better model architecture can help significantly correct and optimize RNNs. Therefore, Hochreitcr and Schmidhuber proposed the LSTM model. To address the issues exposed in RNN models, long lags are added to the LSTM model.

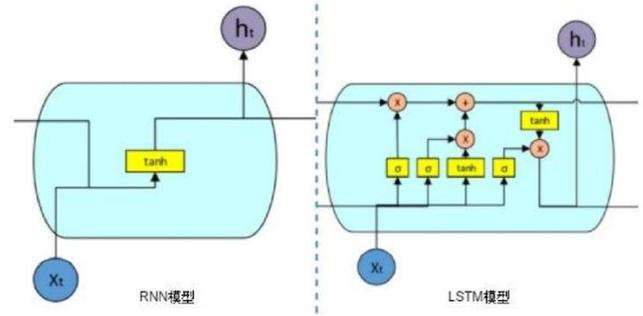

Fig.3 Comparison chart between RNN and LSTM.

As can be seen from the picture on the left, the ht hidden layer state of the RNN model is affected by the previous h (t-1). Through the action of the activation function, the tanh function is generally used as the activation function in RNN. The LSTM is much more complicated. LSTM deepens the internal operation of the hidden layer and sets up three control valves to complete the classification, selection and forgetting process of incoming data. The valve here refers to a combination operation, and its operation process is a sigmoid activation function and a pointwise operation. Ct will also accept the data of the previous period, that is, C (t-1). The forgetting ratio of the previous period's data is controlled by zf. zf and C (t-1) have the same dimensions, and their sigmoid activation method is reflected by the corresponding product of the matrix weight. . The mathematical expression is:

$$f_t = \sigma\left(W_f h_{t-1} + U_f x_t + b_f\right) \# \text{(1)}$$

The activation function is a sigmoid function, and the final output of ft is the forgetting probability.

After C(t-1) or before Ct, the new data input is represented by z, its input proportion is controlled by the control valve zi, and its matrix product is the new added weight.

### B. Theoretical Framework for Trading Strategy Design

This article first conducts a comparative analysis of the principles of three quantitative strategy modeling methods, RNN, LSTM-GRU, and SVM, and then conducts actual testing.

Principle analysis shows that SVM, as a multivariate classification algorithm, inevitably requires the assistance of kernel functions when mapping multi-factor stock data. However, the mapping principle of kernel functions will significantly increase the modeling time and reduce the accuracy when the number of factors increases significantly. ; RNN is a classic stock investment neural network algorithm, but when the time span is long and the number of neurons increases, there will be gradient disappearance and gradient explosion problems, which will affect the modeling effect. The reason why the LSTM algorithm is more suitable for stock investment modeling is that it improves the calculation method of the hidden layer gradient and uses key control valves to improve the disappearance and explosion problems of the gradient. Now the gradient is controlled within a reasonable range, and the input and Output valves can also improve forecasting of long time series. The effectiveness evaluation of the three algorithms is first based on the AUC of a single prediction, and then analyzes the accuracy and stability of multiple prediction AUCs, as well as the time complexity issues that must be considered in machine learning. Choose the best LSTM algorithm. However, the LSTM algorithm also has its own problems, such as its high time complexity, and there is no machine learning algorithm that can handle all data modeling. This article refers to foreign literature in the past two years and uses the GRU model used by domestic scholars just a few months ago to improve it. LSTM model, through multiple AUC prediction tests, it was found that the time complexity of the LSTM-GRU model has been greatly improved. The use of the GRU model alone has been confirmed by academic literature to be less accurate than the LSTM model in text processing, image recognition, and financial index prediction. Therefore, constructing a composite LSTM-GRU model based on GRU is a reasonable research direction. Then, the evaluation is carried out based on the indicator system of the real trading platform (strategy return, strategy annualized return, excess return, benchmark return, alpha, beta, Sharpe ratio, maximum drawdown strategy volatility, and benchmark volatility). The results found that the LSTM-GRU algorithm significantly outperformed the CSI 300 Index, which is often referenced by passive index stock funds, in the long term. At the same time, indicators such as maximum drawdown, beta, alpha, and volatility performed better. Therefore, the theoretical framework of this article consists of three parts, namely data input and cleaning, model modeling, and final portfolio effect evaluation. In other words, the theoretical framework of stock selection strategy needs the following support: as input data, it can meet the modeling requirements; model construction The model can adapt to the actual training of China's stock market; the final and perfect evaluation system can conduct reasonable and effective evaluations of different methods.

## IV. Experiments

### A. Data selection

Based on the consideration of medium and long-term investment plans, this article selects the stock data of the A-share CSI 300 Index from January 1, 2008 to January 1, 2020, including market indicators, financial statement indicators, technical indicators, etc. Take January 1, 2008 to January 1, 2011 as the closed learning period of machine learning, and then conduct rolling backtest training.

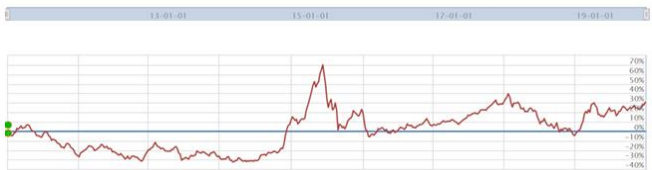

Fig.4 CSI 300 Index earnings chart from 2011 to 2020.

The selection of the CSI 300 Index for empirical research data is mainly based on the following considerations. First, the CSI 300 Index is used by quite a few index funds, so it is easy to represent passive index funds. Secondly, the algorithm process of machine learning is relatively complex. Every time a hidden layer is added, the time may increase by nearly half under the CPU computing power. Due to the limit of computing power, the stock data of CSI 300 is used, which not only ensures sufficient. The number of stocks takes into account efficiency.

### B. Experimental results

Use Krcas's Sequential model to build the network. First build an input layer as input. Then establish a layer of LSMT with 256 neurons, activation function relu, and learning rate 0.0001. In order to prevent overfitting, a layer of Dropout is added later, and the dropout rate parameter is 0.2. Then establish a layer of Dense with 32 neurons and activation function relu. Finally, a layer of Dense is established as the output layer, with a neuron of 1 and an activation function of linear. The model uses compile for model training configuration, set the loss function loss to mse, the optimizer optimizer to adam, and the evaluation indicator metrics to accuracy. Use the ft function to train the model on the training set, epochs=20.

Common evaluation standard indicators are used: mean absolute percentage error function (MAPE), mean absolute error function (MSE) and root mean square error function (RMSE) to evaluate the performance of the model in predicting fund prices. MAPE is a percentage value that measures the relative size of the forecast error as a percentage. MAE stands for mean absolute error, which measures the difference between predicted values and actual observed values. RMSE reflects the absolute error between the predicted value and the true value.

The prediction results of A-share Shanghai and Shenzhen stock data are shown in Figure 5. Among them, the rectangular box legend represents the true value of the fund's closing price, the abscissa represents the number of test sets, and the ordinate represents the fund price.

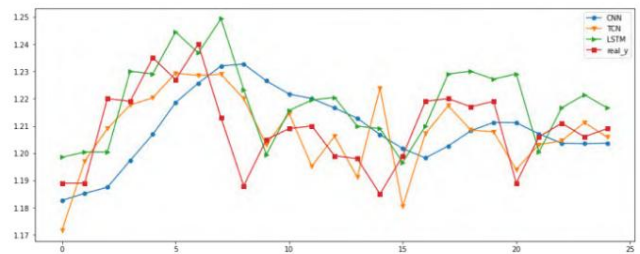

Fig.5 Forecast results chart of A-share Shanghai and Shenzhen stock data.

Comparative experiments were conducted between the LSMT model and CNN and TCN on funds. LSTM has a memory effect on long-term dependencies, can learn effective features from financial sequences, and can correctly predict future development directions. It can be seen that the LSTM model proposed in this article has obvious advantages in predicting stock trading strategies.

## V. Conclusion

This article explores its application potential in the field of stock trading based on the deep learning algorithm of long short-term memory network (LSTM) and gated recurrent unit (GRU). Through empirical research and experimental verification on the A-share CSI 300 Index, we have drawn the following conclusions: First, LSTM and

GRU, as deep learning models, have performed well in stock market prediction and trading strategy design. Effect. Compared with traditional RNN models, LSTM and GRU effectively solve the problems of gradient disappearance and explosion by introducing gating mechanisms and long-short-term memory capabilities, improving the accuracy and stability of the model. Secondly, through comparative experiments, this paper found that the LSTM model has obvious advantages in stock price prediction, especially its ability to learn long-term dependencies and sequence features. Compared with traditional machine learning algorithms such as SVM, the LSTM model shows better performance in both modeling effect and prediction accuracy. In addition, this article also combines comparative experiments between deep learning models and traditional technical analysis methods. The results show that the deep learning model has significant advantages in predicting stock trading strategies. Through empirical research on the A-share CSI 300 Index, we verified the effectiveness and feasibility of the LSTM model in stock market prediction and trading strategy design. Finally, we proposed a theoretical framework for stock trading strategy design based on the LSTM-GRU deep learning model and conducted experimental verification. Experimental results show that this framework can significantly outperform passive index stock funds in the long term, while having better risk control capabilities and stability.